\documentclass[12pt]{article}
  \usepackage{amsfonts}
  \usepackage{amsmath}
\usepackage{amssymb}
\usepackage{amscd}
\usepackage[dvips]{graphicx}
\begin{document}

~~
\bigskip
\bigskip
\begin{center}
{\Large {\bf{{{Classical mechanics  of many particles defined on
canonically deformed nonrelativistic space-time}}}}}
\end{center}
\bigskip
\bigskip
\bigskip
\begin{center}
{{\large ${\rm {Marcin\;Daszkiewicz^{1}}}$, \large ${\rm {Cezary\;
 J.\;Walczyk^{2}}}$}}
\end{center}
\bigskip
\begin{center}
\bigskip

{ ${\rm{~^{1}Institute\; of\; Theoretical\; Physics}}$}

{ ${\rm{ University\; of\; Wroclaw\; pl.\; Maxa\; Borna\; 9,\;
50-206\; Wroclaw,\; Poland}}$}

{ ${\rm{ e-mail:\; marcin@ift.uni.wroc.pl}}$}

\bigskip

{ ${\rm{~^{2}Department\; of\; Physics}}$}

{ ${\rm{ University\; of\; Bialystok,\; ul.\; Lipowa\; 41,\;
15-424\;Bialystok,\; Poland}}$}

{ ${\rm{ e-mail:\; c.walczyk@alpha.uwb.edu.pl}}$}

\end{center}
\bigskip
\bigskip
\bigskip
\bigskip
\bigskip
\bigskip
\bigskip
\bigskip
\bigskip
\begin{abstract}
We provide the classical mechanics  of many particles moving in
canonically twist-deformed  space-time. In particular, we consider
two examples of such noncommutative systems - the set of $N$
particles moving in gravitational field as well as the system of $N$
interacting harmonic oscillators.
\end{abstract}
\bigskip
\bigskip
\bigskip
\bigskip
\eject

\section{{{Introduction}}}

The idea to use noncommutative coordinates is quite old - it goes
back to Heisenberg and was firstly formalized by Snyder in
\cite{snyder}. Recently, however,  there were  found new formal
arguments based mainly on Quantum Gravity \cite{2}, \cite{2a} and
String Theory models \cite{recent}, \cite{string1}, indicating that
space-time at Planck scale  should be noncommutative, i.e. it should
have a quantum nature. On the other side, the main reason for such
considerations follows from many phenomenological considerations,
which state that relativistic space-time symmetries should be
modified (deformed) at Planck scale, while  the classical Poincare
invariance still remains valid at larger distances
\cite{1a}-\cite{1d}.

It is well-known that  a proper modification of the Poincare and
Galilei Hopf algebras can be realized in the framework of Quantum
Groups \cite{qg1}, \cite{qg3}. Hence, in accordance with the
Hopf-algebraic classification  of all deformations of relativistic
and nonrelativistic symmetries (see \cite{class1}, \cite{class2}),
one can distinguish two
  simplest quantum spaces. First of them
corresponds to the well-known canonical type of noncommutativity
\begin{equation}
[\;{ x}_{\mu},{ x}_{\nu}\;] =
i\theta_{\mu\nu}\;,
\label{wielkaslawia}
\end{equation}
with antisymmetric constant tensor $\theta^{\mu\nu}$. Its
relativistic and nonrelativistic Hopf-algebraic realizations have
been discovered  with the use of  twist procedure (see \cite{twist})
of classical Poincare \cite{oeckl}, \cite{chi} and Galilei
\cite{daszgali}, \cite{daszdual} Hopf structures respectively.\\
The second class of  deformations  introduces the Lie-algebraic type
of space-time noncommutativity
\begin{equation}
[\;{ x}_{\mu},{ x}_{\nu}\;] = i\theta_{\mu\nu}^{\rho}{ x}_{\rho}\;,
\label{betanoncomm1}
\end{equation}
with particularly chosen constant coefficients
$\theta_{\mu\nu}^{\rho}$. The examples of corresponding Poincare
quantum algebras have been introduced in \cite{lie1}, \cite{lie2},
while the suitable Galilei algebras  - in \cite{kappaG},
\cite{daszgali} and \cite{daszdual}.

Recently, there appeared a lot of papers dealing with classical
(\cite{deri}-\cite{daszwal}) and quantum (\cite{qm1}-\cite{oscy})
mechanics, Doubly Special Relativity frameworks (\cite{dsr1a},
\cite{dsr1b}), statistical physics (\cite{maggiore}, \cite{rama})
and field theoretical models (see e.g. \cite{przeglad}), defined on
quantum
 space-times (\ref{wielkaslawia}),
 (\ref{betanoncomm1})\footnote{For earlier studies see \cite{lukiluk1} and
\cite{lukiluk2}.}. Particulary, there was
 investigated the impact of the mentioned
above   deformations  on dynamics of basic classical and quantum
systems. Consequently, in  papers \cite{romero}, \cite{romero1}, the
authors considered  classical particle moving in central
gravitational field defined on canonically deformed space-time
(\ref{wielkaslawia}). They  demonstrated, that in such a case there
is generated Coriolis force acting additionally on the moving
particle. Besides, in articles \cite{lodzianieosc}, \cite{romero}
and \cite{oscy} there was analyzed classical and quantum oscillator
model formulated on canonically and Lie-algebraically deformed
space-time respectively. Particulary, there has been found its
deformed energy spectrum as well as the corresponding equation of
motion. Interesting results have been also obtained in two papers
\cite{qm1}, \cite{qm2} concerning the  hydrogen atom model defined
on spaces (\ref{wielkaslawia})  and (\ref{betanoncomm1}). Besides,
it should be noted that  there appeared article \cite{toporzelek},
which provides the link between Pioneer anomaly phenomena
\cite{piophen} and classical mechanics defined on $\kappa$-Galilei
quantum space. Preciously,  there has been demonstrated that
additional force term acting on moving satellite can be identified
with the force generated  by space-time noncommutativity. The value
of deformation parameter $\kappa$ has been fixed  by comparison of
obtained theoretical results with observational data.

Unfortunately, in all mentioned above articles there were analyzed
only the one-particle relativistic and nonrelativistic dynamics in
the field of forces. Here, we extend a such kind of investigations
to the classical mechanics of many particles, which move in the
modified canonically deformed space-time\footnote{$x_0 =
ct$.},\footnote{It should be noted that a such modification of relation (\ref{wielkaslawia}) (blind in $a$, $b$ indieces) is in accordance with the formal arguments proposed in \cite{fiore}. Preciously, the relations (3) are constructed with adopt so-called braided tensor algebra procedure, dictated by structure of quantum $R$-matrix for canonical deformation [10], [11]. Such a choice is compatibile with Leibnitz rules for quantum algebra given by deformed coproduct (\ref{dlww3v})-(\ref{zadruzny}).}
\begin{equation}
[\,t,x^i_a\,] = 0\;\;\;,\;\;\;[\,x^i_a,x^j_b\,] =
i\theta^{ij}\;\;\;;\;\;\;i, j = 1,2, 3\;, \label{canamm}
\end{equation}
with  indices $a, b = 1,2, \ldots ,N$ labeling the particle.
Further, we indicate that as in the case of one-particle system
there appeared additional force terms generated by space-time
noncommutativity. Of course, for $N=1$  our results become the same
as the results obtained in \cite{romero}, \cite{lodzianieosc}.

The motivations for present studies are manyfold. First of all we
extend in natural way the  results for one-particle model to the
many-particle system. Secondly, such investigations permit to
analyze the deformations of wide class of physical models, for
example, one can applied the presented results to the studies on two
deformed systems considered in Sect. 4. Finally, it gives a starting
point for the construction of nonrelativistic quantum mechanics for
many particles defined on modified space-time (\ref{canamm}).

The paper is organized as follows. In Sect. 2 we recall basic facts
concerning the twisted canonically deformed Galilei Hopf algebra
$\,{\mathcal U}_{\theta}(\mathcal{G})$ associated with space-time
noncommutativity (\ref{wielkaslawia}) for $\theta_{0i} = 0$ and
$\theta_{ij} \ne 0$. In Sect. 3 we provide the classical
many-particle model defined on modified canonical space-time
(\ref{canamm}). Section four includes two prominent\footnote{Their
classical (undeformed) versions are often discussed in the
literature, see e.g. \cite{sym}.} examples of such deformed systems
- the model of $N$ particles moving in central gravitational field
as well as  the system of $N$ coupling harmonic oscillators. The
final remarks are presented in the last section.

\section{{{Twisted Galilei Hopf algebra and corresponding canonically  deformed space-time}}}

In accordance with Drinfeld twist procedure [14], [10], [11], the algebraic sector of arbitrary twisted
 Hopf algebra $\,\mathcal{U}({A})$  remains undeformed, while the coproducts and antipodes transform as follows
\begin{equation}
 \Delta _{0}(a) \longrightarrow \Delta(a) = \mathcal{F}\circ
\,\Delta _{0}(a)\,\circ \mathcal{F}^{-1}\;\;\;,\;\;\;
S(a) =u\,S_{0}(a)\,u^{-1}\;,\label{fs}
\end{equation}
with $\Delta _{0}(a) = a \otimes 1 + 1 \otimes a$, $S_0(a) = -a$ and
$u=\sum f_{(1)}S_0(f_{(2)})$ (we use Sweedler's notation
$\mathcal{F}_{\cdot }=\sum f_{(1)}\otimes f_{(2)}$).
Besides, present in the above formula  twist factor
$\mathcal{F} \in {\mathcal U}(A) \otimes {\mathcal U}(A)$
satisfies  the classical cocycle condition
\begin{equation}
{\mathcal F}_{12} \cdot(\Delta_{0} \otimes 1) ~{\cal
F} = {\mathcal F}_{23} \cdot(1\otimes \Delta_{0})
~{\mathcal F}\;, \label{cocyclef}
\end{equation}
and the normalization condition
\begin{equation}
(\epsilon \otimes 1)~{\cal F}= (1 \otimes
\epsilon)~{\cal F} = 1\;, \label{normalizationhh}
\end{equation}
with ${\cal F}_{12} = {\cal F}\otimes 1$ and
${\cal F}_{23} = 1 \otimes {\cal F}$.

Consequently, in the case of the canonically
deformed Galilei Hopf algebra $\,{\mathcal U}_{\theta}(\mathcal{G})$
provided in \cite{daszgali}, we have
\begin{equation}
\mathcal{F}_{\theta} = \exp \,\left(-\frac{i}{4}\theta^{ij}\Pi_{i}
\wedge \Pi_{j} \right)\;,\label{factor}
\end{equation}
and, in accordance with (\ref{fs}), we get the following algebraic\footnote{The symbols $K_{ij}$, $V_i$ and $\Pi_\mu$
denote rotations, boosts  and space-time translation generators
respectively.}
\begin{eqnarray}
&&\left[\, K_{ij},K_{kl}\,\right] =i\left( \delta
_{il}\,K_{jk}-\delta
_{jl}\,K_{ik}+\delta _{jk}K_{il}-\delta _{ik}K_{jl}\right) \;,  \notag \\
&~~&  \cr &&\left[\, K_{ij},V_{k}\,\right] =i\left( \delta
_{jk}\,V_i-\delta _{ik}\,V_j\right)\;\; \;, \;\;\;\left[
\,K_{ij},\Pi_{k }\,\right] =i\left( \delta _{j k }\,\Pi_{i }-\delta
_{ik }\,\Pi_{j }\right) \;, \label{nnnga}
\\
&~~&  \cr &&\left[ \,K_{ij},\Pi_{0 }\,\right] =\left[
\,V_i,V_j\,\right] = \left[ \,V_i,\Pi_{j }\,\right]
=0\;\;\;,\;\;\;\left[ \,V_i,\Pi_{0 }\,\right]
=-i\Pi_i\;\;\;,\;\;\;\left[ \,\Pi_{\mu },\Pi_{\nu }\,\right] =
0\;,\nonumber
\end{eqnarray}
and the coalgebraic
\begin{eqnarray}
&&\Delta_\theta(\Pi_\rho)=\Delta_0(\Pi_\rho)\;\;\;,\;\;\;
\Delta _{\theta }(V_i) =\Delta _{0}(V_i)\;, \label{dlww3v}\\
&~~~&  \cr \Delta _{\theta }(K_{ij})
&=&\Delta _{0}(K_{ij})-%
\theta ^{k l }[(\delta_{k i}\Pi_{j }-\delta_{k j
}\,\Pi_{i})\otimes \Pi_{l }\\
&&\qquad\qquad\qquad\qquad\qquad+\Pi_{k}\otimes (\delta_{l
i}\Pi_{j}-\delta_{l j}\Pi_{i})]\;,\label{zadruzny}
\end{eqnarray}
sectors.

It is well-known (see e.g. \cite{chi}) that the deformed space-time
corresponding to the Hopf algebra $\,{\mathcal
U}_{\theta}(\mathcal{G})$ is defined as the quantum representation
space (Hopf module),  with action of the deformed symmetry
generators satisfying suitably deformed Leibnitz rules.  The action
of Galilei group $\,{\mathcal U}_{\theta}(\mathcal{G})$ on a Hopf
module of functions depending on space-time coordinates $(t,x_i)$ is
given by
\begin{eqnarray}
&&\Pi_{0}\rhd
f(t,\overline{x})=i{\partial_t}f(t,\overline{x})\;\;\;,\;\;\;
\Pi_{i}\rhd f(t,\overline{x})=i{\partial_i}f(t,\overline{x})\;,
\label{a1}\\
&~~&  \cr &&K_{ij}\rhd f(t,\overline{x}) =i\left( x_{i }{\partial_j}
-x_{j }{\partial_i} \right) f(t,\overline{x})\;\;\;,\;\;\; V_i\rhd
f(t,\overline{x}) =it{\partial_i}
\,f(t,\overline{x})\;,~~~~~~~~\label{dsf}
\end{eqnarray}
while the $\star$-multiplication of arbitrary two functions  is
defined as follows
\begin{equation}
f(t,\overline{x})\star_{{\theta}} g(t,\overline{x}):=
\omega\circ\left(
 \mathcal{F}_{\theta}^{-1}\rhd  f(t,\overline{x})\otimes g(t,\overline{x})\right)
 \;\;\;;\;\;\;\omega\circ\left( a\otimes b\right) = a\cdot b\;.
\label{star}
\end{equation}
Consequently, in the representation (\ref{a1}), (\ref{dsf}) the
twist factor (\ref{factor}), and the corresponding nonrelativistic
space-time  take respectively the forms
\begin{equation}
\mathcal{F}_{{\theta}}=  {\rm \exp}
\,\left(\frac{i}{4}{{\theta^{ij}}}\,\partial_{i}\wedge {\partial_j}
\right)\;, \label{swar1}
\end{equation}
and
\begin{equation}
[\,t,x_i\,]_{{\star}_{{\theta }}} =
0\;\;\;,\;\;\;[\,x_i,x_j\,]_{{\star}_{{\theta }}} =
 i\theta^{ij}
 \;,
\label{stcom1}
\end{equation}
 with $[\, a,b\,] _{\star_{{\theta}}}:=a{\star_{{\theta}}} b -
b{\star_{{\theta}}}a$. Obviously, for  deformation parameter
$\theta_{ij}$ approaching zero the above quantum space becomes the
classical one.

\section{{{Classical mechanics of  many particles moving in canonically deformed space-time}}}

In this section we provide the classical model of $N$
nonrelativistic particles moving in   canonically deformed
 space-time (\ref{canamm}), which for $N=1$ reproduces commutation relations
 (\ref{stcom1}). As it was mentioned in
Introduction,  similar constructions have been performed in the case
of one-particle system in the series of papers \cite{romero},
\cite{romero1} and \cite{lodzianieosc}.

In a first step of our investigation we start with the following
phase space\footnote{We use the correspondence relation $\{\;a,b\;\}
= \frac{1}{i}[\;\hat{a},\hat{b}\;]$  $(\hbar = 1)$.}
\begin{equation}
\{\,x^i_a,x^j_b\,\} = \theta^{ij}\;, \label{cana}
\end{equation}
\begin{equation}
\{\,p^i_a,p^j_b\,\}=0\;\;\;,\;\;\;\{\,x^i_a,p^j_b\,\}=\delta^{ij}\delta_{ab}\;.
\label{can}
\end{equation}
which  satisfies the Jacobi identity and for $N=1$  becomes the same
as phase space provided in
\cite{romero}.\\
Next, following \cite{sym},  for  arbitrary two functions
$F(\zeta^A)$ and $G(\zeta^A)$ we define  Poisson bracket as follows
\begin{equation}
 \{\,F,G\,\}=\sum_{A,B = 1}^{6N}\{\,\zeta^A,\zeta^B\,\}\
\frac{\partial F}{\partial \zeta^A}\ \frac{\partial G}{\partial
\zeta^B}\;,\label{motion1}
\end{equation}
with $\zeta^A = (x^i_a,p^i_a)$.\\
In terms of the above structure and given  Hamiltonian $H =
H(\zeta^A)$ one can write the equations of motion as
\begin{equation}
\dot{\zeta}^A=\{\,\zeta^A,H\,\}\;\;\;;\;\;\;\dot{\zeta}^A:=\frac{d\zeta^A}{dt}\;.\label{motion2}
\end{equation}
Moreover, in general case, i.e. for any function $F$ depending on
$\zeta^A$, we have
\begin{equation}
\dot{F}=\{\,F,H\,\}\;.\label{motion1}
\end{equation}

Let us now introduce  the standard Hamiltonian function describing
the set of $N$ particles
\begin{eqnarray}
H(\vec{p}_1,\ldots,\vec{p}_N,\vec{r}_1,\ldots,\vec{r}_N) =
\sum_{a=1}^{N}\frac{\vec{p}_a^{\,2}}{2m_a} +
V(\vec{r}_1,\ldots,\vec{r}_N)\;,\label{ham}
\end{eqnarray}
with $\vec{p}_a = [\,p^1_a,p^2_a,p^3_a\,]$ and $\vec{r}_a =
[\,x^1_a,x^2_a,x^3_a\,]$. Then, in accordance with (\ref{motion2}),
we get the following $2N$ equations of motion
\begin{equation}
\dot{x}^i_a =  \frac{p^i_a}{m_a}  + \sum_{b = 1}^{N} \sum_{j =
1}^{3} \theta^{ij}\,\frac{\partial V}{\partial x^j_b} \;\;\;,\;\;\;
 \dot{p}^i_a =  -\frac{\partial V}{\partial
x^i_a}\;, \label{ham2}
\end{equation}
which lead to $N$  Newton equations
\begin{eqnarray}
m_a\ddot{x}^i_a  = -\frac{\partial V}{\partial x^i_a} + m_a\sum_{b,
c = 1}^{N}\sum_{j,k = 1}^{3} \theta^{ij}\,\frac{\partial^2
V}{\partial x^j_b\partial
 x^k_c}\dot{x}^k_c
\;. \label{cannewton}
\end{eqnarray}
It should be noted that in the case of one-particle system $(N=1)$
the above equation  becomes the same as the one derived in \cite{romero}. \\
Further, we  consider the multiparticle potential function  given by
\begin{eqnarray}
V(\vec{r}_1,\ldots,\vec{r}_N) = \sum_{a = 1}^{N} V_a^{\rm
out}(|\vec{r}_a|) + \frac{1}{2}\sum_{a=1}^N\sum_{a \ne b}
V_{ab}^{\rm inn}(|\vec{r}_a- \vec{r}_b|)\;, \label{pot}
\end{eqnarray}
where in the above formula  symbol $V_a^{\rm out}$ denotes the
"outer" potential   acting on $a$-th particle, and $V_{ab}^{\rm inn}
= V_{ba}^{\rm inn}$ corresponds to the so-called "inner" potential
describing interactions between particles $a$ and $b$. Besides, we
have
\begin{equation}
r_a = |\vec{r}_a| = \sqrt{\sum_{i=1}^3 x_a^i\cdot
x_a^i}\;\;\;\;\;,\;\;\;\;\;r_{ab} = |\vec{r}_a- \vec{r}_b|\;.
\label{dlugosci}
\end{equation}
One can  check, that in the case of hamiltonian function (\ref{ham})
the equations of motion (\ref{ham2}) look as follows
\begin{eqnarray}
\dot{x}_a^i&=&\frac{p_a^i}{m_a}+\sum_{b=1}^N\sum_{j=1}^3
\theta^{ij}\frac{1}{r_b}\frac{\partial V_b^{out}}
{\partial r_b}x_b^j\;,\\
\dot{p}_a^i &=&-\frac{1}{r_a}\frac{\partial V_a^{out}}{\partial
r_a}x_a^i-\sum_{b\neq a}\frac{1}{r_{ab}}\frac{\partial
V_{ab}^{inn}}{\partial
r_{ab}}\left(x_a^i-x_b^i\right)\;,\label{hamg}
\end{eqnarray}
while the Newton equations (\ref{cannewton})  take the form
\begin{eqnarray}
&&m_a\ddot{x}^i_a  = -\frac{{x}^i_a}{r_a}\frac{\partial
V_a^{\rm{out}}}{\partial r_a} -\sum_{b \ne a}
\frac{(x_a^i-x_b^i)}{r_{ab}}\frac{\partial
V_{ab}^{\rm{inn}}}{\partial r_{ab}}\;
+~~~~~~~~~~~~~~~~~~~~~~~~~~~~~~~~~~~~~~~~ \cr
&&~~~~~~~~~~~~~~~~~~~~~~~~~~~~~~-\,m_a\sum_{b =
1}^{N}\sum_{j,k=1}^3\epsilon^{ikj}\left(\dot{\Omega}_k^b(x,\dot{x})
x^j_b+{\Omega}_k^b(x) \dot{x}^j_b \right)\;,\label{szukalski}
\end{eqnarray}
with
\begin{eqnarray}
\theta^{ij} &=& \sum_{k=1}^3\epsilon^{ijk}\theta_k\;,\cr
 \Omega_i^a(x) &=&
\frac{1}{r_a}\frac{\partial V_a^{\rm{out}}}{\partial
r_a}\theta_i\;,\label{coef}\\
\dot{\Omega}_i^a(x,\dot{x}) &=&
\frac{\theta_i}{r_a^2}\left(\frac{\partial^2
V_{a}^{\rm{out}}}{\partial r_{a}^2} - \frac{1}{r_a}\frac{\partial
V_{a}^{\rm{out}}}{\partial r_{a}} \right)\vec{x}_a\cdot
\dot{\vec{x}}_a \;. \nonumber
\end{eqnarray}
Finally, it should be  noted that the following  position and
momentum dependent function
\begin{equation}
L_\theta(x,p)= \sum_{a=1}^N\sum_{i,j=1}^3 \theta^{ij} x_a^i
p_a^j+\frac{1}{2}\sum_{i,j,k=1}^3
\theta^{ij}P^j\theta^{ik}P^k\;\;\;;\;\;\;  P^j=\sum_{a=1}^N
p_a^j\;,\label{wiel}
\end{equation}
plays the role of  constant of motion, i.e. it satisfies the
equation
\begin{equation}
\dot{L}_\theta =  \{\,L_\theta,H\,\}= 0\;,\label{ewolucja}
\end{equation}
and it transforms the phase space variables as follows\footnote{One can notice, that for  $\theta^{ij} = \sum_{k=1}^3\epsilon^{ijk}\theta_k$ the function $L_\theta$  generates  rotations in $\theta_i$-directions.}
\begin{eqnarray}
\{\,{x}_a^i,L_\theta\,\}= \sum_{j=1}^{3}\theta^{ij}x_a^j\;\;\;,\;\;\;
\{\,{p}_a^i,L_\theta\,\} = - \sum_{j=1}^{3}\theta^{ij}p_a^j\;.\label{trans2}
\end{eqnarray}
Besides, one should  observe that the standard  total momentum
\begin{equation}
\vec{P} =\sum_{a=1}^{N} \vec{p}_a\;,\label{totalmomentum}
\end{equation}
 as well as the standard total angular momentum
\begin{equation}
\vec{L} = \sum_{a=1}^{N}\vec{L}_a\;\;\;;\;\;\; \vec{L}_a=
\sum_{a=1}^{N} \vec{r}_a \times \vec{p}_a\;,\label{totalangmomentum}
\end{equation}
are not conserved in time due to the presence of "outer" potential $
V_a^{\rm out}$, i.e.
\begin{equation}
\dot{\vec{P}} =  \{\,\vec{P},H\,\}= -
\sum_{a=1}^{N}\frac{\vec{x}_a}{r_a}\frac{\partial V_a^{\rm
out}}{\partial r_a}\;,\label{dendryd}
\end{equation}
and
\begin{equation}
\dot{\vec{L}} =
\{\,\vec{L},H\,\}=\sum_{a,b=1}^{N}\frac{1}{r_b}\frac{\partial
V_b^{\rm out}}{\partial
r_b}\vec{p}_a\times\left(\vec{\theta}\times\vec{x}_b\right)\;,
\label{dendryd1}
\end{equation}
respectively, with $\vec{\theta} =
[\,\theta_1,\theta_2,\theta_3\,]$.\\
Obviously, for deformation parameter $\theta^{ij}\, (\theta^k)$
approaching zero the above model becomes undeformed, i.e. we recover
the ordinary Newton mechanics for set of $N$ particles  \cite{sym}.

\section{{{Two examples of canonically deformed many-particle models}}}

In this section we discuss  two selected examples (see e.g.
\cite{sym}) of many-particle systems defined on noncommutative
space-time (\ref{canamm}), namely the  system of $N$ particles
moving in gravitational field, and the set of $N$  interacting
harmonic oscillators.

\subsection{{{The  system of $N$ particles moving in
gravitational field}}}

Let  us  consider the system of $N$ particles moving in the presence
of mass $M$ located in the origin of the coordinate system. Then,
the "outer" potential takes the form
\begin{eqnarray}
 V_a^{\rm
out}(|\vec{r}_a|)  = -\frac{GMm_a}{|\vec{r}_a|}\;, \label{nnnnn}
\end{eqnarray}
while the "inner" potentials look as follows
\begin{eqnarray}
V_{ab}^{\rm inn}(|\vec{r}_a- \vec{r}_b|) =-\frac{Gm_a
m_b}{|\vec{r}_a - \vec{r}_b|} \;. \label{pot1}
\end{eqnarray}
One can  check that for potential functions (\ref{nnnnn}) and
(\ref{pot1}) we get the following canonical equation of motion
\begin{eqnarray}
\dot{x}_a^i&=&\frac{p_a^i}{m_a}+\sum_{b=1}^N\sum_{j=1}^3
\theta^{ij}\frac{G M m_b}{r_b^3}x_b^j\;,\\
\dot{p}_a^i &=&-\frac{G M m_a}{r_a^3}x_a^i-\sum_{b\neq a}\frac{G m_a
m_b}{r_{ab}^3}\left(x_a^i-x_b^i\right)\;,\label{gravcanon}
\end{eqnarray}
leading to the set of Newton equations
\begin{equation*}
m_a\ddot{x}_a^i=-Gm_a \left(\frac{x_a^i}{r_a}\frac{M}{ r_a^2}
+\sum_{b \ne a} \frac{(x_a^i-x_b^i)}{r_{ab}}\frac{m_b}{ r_{ab}^2}
\right)+~~~~~~~~~~~~~~~~~~~~~~~~~~~~~~~~~~~~
\end{equation*}
\begin{equation}
~~~~~~~~~~~~~~~~~~~~~~~~~-m_a\sum_{b=1}^N\sum_{j,k=1}^3\epsilon^{ikj}\left(\dot{{\Omega}}_k^b(x,\dot{x})
x_b^j+{\Omega}_k^b(x) \dot{x}_b^j\right)\;,\label{czartwor}
\end{equation}
with
\begin{equation}
{{\Omega}}_k^b({x})=G\frac{Mm_b}{r_b^3}\theta_k \;,\label{perunwit}
\end{equation}
and
\begin{equation}
\dot{{\Omega}}_k^b(x,\dot{x})=-3G\theta_k\frac{Mm_b}{r_b^5}\vec{x}_b\dot{\vec{x}}_b\;.\label{swarzyca}
\end{equation}
Obviously, 
for mass $M$  equal zero, i.e. for vanishing external potential
$V_a^{\rm out}$, the equation of motion (\ref{czartwor})  takes the
standard form
\begin{equation}
m_a\ddot{x}_a^i=-\sum_{b \ne a}
\frac{(x_a^i-x_b^i)}{r_{ab}}\frac{Gm_am_b}{ r_{ab}^2}
\;.\label{czartwor1}
\end{equation}
Besides, one can observe that for $N=1$ the above system reproduces
the well-known  model for single particle moving in gravitational
field \cite{sym}.

In order to analyze the above system in terms of commutative
variables  $(t,x_{\rm com})$, by analogy to the star multiplication
(\ref{star}), one should replace the product of two  arbitrary
functions defined on quantum space (\ref{canamm}), by the following
${\hat\star}_{{\theta}}$-product\footnote{One can derive the
commutation relations (\ref{canamm}) using definition (\ref{stcom1})
with inserted ${\hat\star}_{{\theta}}$-product instead
${\star}_{{\theta}}$-multiplication (see also footnote 6).} (see \cite{mech})
\begin{equation}
f(t,\overline{x})\cdot g(t,\overline{x}) \longrightarrow
f(t,\overline{x}_{{\rm com}})\,{\hat\star}_{{\theta}}\,
g(t,\overline{x}_{{\rm com}}):= \omega\circ\left(
 \mathcal{O}_{\theta}\rhd  f(t,\overline{x})\otimes g(t,\overline{x})\right)\;, \label{genstar}
\end{equation}
with
\begin{equation}
\mathcal{O}_{{\theta}}=  {\rm \exp}
\,\left(\frac{1}{4}\sum_{a,b=1}^{N}\sum_{i,j=1}^3{{\theta_{ij}}}\,\partial^{i}_{a\,\rm
com}\wedge
\partial^{j}_{b\,\rm
com} \right)\;. \label{genswar1}
\end{equation}
Then, we get
\begin{equation*}
m_a\ddot{x}_{a\,\rm com}^i=-Gm_a
\left(M{x_a^i}\,{\hat\star}_{{\theta}} \left(\frac{1}{r_{a\,\rm
com}}\right){\hat\star}_{{\theta}}\left(\frac{1}{r_{a\,\rm
com}}\right){\hat\star}_{{\theta}}\left(\frac{1}{r_{a\,\rm
com}}\right)\right. +~~~~~~~~~~~~~~~~~~~
\end{equation*}
\begin{equation}
\left. +\sum_{b \ne a}m_b {\left(x_{a\,\rm com}^i-x_{b\,\rm
com}^i\right)}\,{\hat\star}_{{\theta}} \left(\frac{1}{r_{ab\,\rm
com}}\right){\hat\star}_{{\theta}} \left(\frac{1}{r_{ab\,\rm
com}}\right){\hat\star}_{{\theta}} \left(\frac{1}{r_{ab\,\rm
com}}\right) \right)+\label{bogorly}
\end{equation}
\begin{equation*}
~~~~~~~~~~~~~~~~~~~~~~~~~~~~-m_a\sum_{b=1}^N\sum_{j,k=1}^3\epsilon^{ikj}\left(\dot{{\Omega}}_k^b(x_{\rm
com},\dot{x}_{\rm com}){\hat\star}_{{\theta}} x_{b\,{\rm
com}}^j+{\Omega}_k^b(x_{\rm com})
{\hat\star}_{{\theta}}\dot{x}_{b\,{\rm com}}^j\right)\;,
\end{equation*}
where
\begin{equation}
{{\Omega}}_k^b({x_{\rm com}})=GMm_b\theta_k\left(\frac{1}{r_{b\,\rm
com}}\right){\hat\star}_{{\theta}}\left(\frac{1}{r_{b\,\rm
com}}\right){\hat\star}_{{\theta}}\left(\frac{1}{r_{b\,\rm
com}}\right) \;,\label{swarzyca120}
\end{equation}
\begin{eqnarray}
\dot{{\Omega}}_k^b(x_{\rm com},\dot{x}_{\rm
com})&=&-3G\theta_kMm_b\left(\frac{1}{r_{b\,\rm
com}}\right){\hat\star}_{{\theta}}\left(\frac{1}{r_{b\,\rm
com}}\right){\hat\star}_{{\theta}}\left(\frac{1}{r_{b\,\rm
com}}\right){\hat\star}_{{\theta}}\cr &{\hat\star}_{{\theta}}&
\left(\frac{1}{r_{b\,\rm
com}}\right){\hat\star}_{{\theta}}\left(\frac{1}{r_{b\,\rm
com}}\right){\hat\star}_{{\theta}}\, \vec{x}_{b\,\rm
com}\,{\hat\star}_{{\theta}}\,\dot{\vec{x}}_{b\,\rm
com}\;,\label{toporzel1000}
\end{eqnarray}
and
\begin{equation}
r_{a\,{\rm com}} = \sqrt{\sum_{i=1}^3x^i_{a\,{\rm
com}}{\hat\star}_{{\theta}}x^i_{a\,{\rm com}}} =
\sqrt{\sum_{i=1}^3x^i_{a\,{\rm com}}\cdot x^i_{a\,{\rm com}}}
\;.\label{rownosc}
\end{equation}
\begin{equation}
x^i_{a\,{\rm com}}{\hat\star}_{{\theta}}\, \dot{x}_{a\,{\rm com}}^i
= x^i_{a\,{\rm com}}\cdot \dot{x}_{a\,{\rm com}}^i \label{rownosc2}
\end{equation}
Further, one can check that linear in deformation parameter
$\theta^k$ corrections, which appear in the  equation
(\ref{bogorly}), look as follows
\begin{equation}
m_a\ddot{x}_{a\,\rm com}^i=-Gm_a \left(\frac{M\,x_{a\,\rm
com}^i}{r^3_{a\,\rm
com}}-\frac{3}{2}\sum_{j=1}^3\theta_{ij}\frac{M\,x_{a\,\rm
com}^j}{r^5_{a\,\rm
com}}+\sum_{b \ne a}\frac{m_b \left(x_{a\,\rm com}^i-x_{b\,\rm
com}^i\right)}{r^3_{ab\,\rm com}} \right)+\label{bogorly2}
\end{equation}
\begin{equation*}
~~~~~~~~~~~~~~~~~~~~~~~~~~~~-m_a\sum_{b=1}^N\sum_{j,k=1}^3\epsilon^{ikj}\left(\dot{{\overline\Omega}}_k^b(x_{\rm
com},\dot{x}_{\rm com})\, x_{b\,{\rm
com}}^j+{{\overline\Omega}}_k^b(x_{\rm com}) \,\dot{x}_{b\,{\rm
com}}^j\right)\;,
\end{equation*}
with
\begin{equation}
{{\overline\Omega}}_k^b({x_{\rm
com}})=\theta_k\frac{GMm_b}{r_{b\,\rm
com}^3}
\;,\label{linearswarzyca120}
\end{equation}
and
\begin{eqnarray}
\dot{{\overline\Omega}}_k^b(x_{\rm com},\dot{x}_{\rm
com})&=&-3\theta_k\frac{GMm_b}{r_{b\,\rm
com}^5}\vec{x}_{b\,\rm com}\cdot\dot{\vec{x}}_{b\,\rm
com}\;,\label{lineartoporzel1000}
\end{eqnarray}
Unfortunately, due to the complicated form of the above Newton
equation, its solution can be studied  by using only  numerical
methods. Such an investigation will be omitted in present article.

\subsection{The system of $N$ coupled harmonic oscillators}

Let us now turn to the second  example of multiparticle system - the
model of $N$ coupled harmonic oscillators. In such a case the
"outer" and "inner" potentials take the form
\begin{eqnarray}
 V_a^{\rm
ext}(|\vec{r}_a|)  =
\frac{m_a\omega_a^2}{2}|\vec{r}_a|^2\;,\label{potosc1}
\end{eqnarray}
and
\begin{eqnarray}
 V_{ab}^{\rm inn}(|\vec{r}_a- \vec{r}_b|) =\frac{\lambda_{ab}^2}{2}
 |\vec{r}_a- \vec{r}_b|^2
\;, \label{potosc2}
\end{eqnarray}
with $\omega_a$ and $\lambda_{ab}$ denoting the frequency and
coupling constant respectively.\\
The corresponding  Newton equation looks as follows
\begin{equation}
m_a\ddot{x}_a^i=-\left(m_a\omega_a^2 + \sum_{b\ne
a}\lambda_{ab}^2\right) x_a^i + \sum_{b\ne a}\lambda_{ab}^2 x_b^i -
m_a
\sum_{b=1}^{N}\sum_{j,k=1}^{3}\epsilon^{ikj}\Omega_{k}^b\dot{x}_b^j\;,\label{czartwor189}
\end{equation}
where
\begin{equation}
\Omega_{i}^a = m_a\omega_a^2\theta_i = {\rm
const.}\;\;\;(\dot{\Omega}_{i}^a =0)\;.\label{zadrugos}
\end{equation}
Of course, in the case of single oscillator model ($N=1$,
$\lambda_{ab}=0$) one recovers the equation of motion proposed in
\cite{romero}.

As an illustration let us consider the case of $m_a = m$, $\omega_a
=\omega$ and $\lambda_{ab} = \lambda$ for $a,b = 1,2,\ldots N$.
Then, it follows from (\ref{czartwor189}) that the corresponding
equation of motion takes the form
\begin{equation}
\ddot{x}_a^i=-\left(\omega^2+\frac{\lambda^2}{m}\sum_{b\neq
a}^N\right)x_a^i+ \frac{\lambda^2}{m}\sum_{b\neq a}^N
x_b^i-m\omega^2 \sum_{b=1}^N \sum_{j,k=1}^3 \epsilon^{ijk}\theta_k
\dot{x}_b^j\;,\label{szukalski201}
\end{equation}
and,   due to linearity of the above formula with respect
$x$-variables, its form remains the same on commutative space-time
as (\ref{szukalski201}).

The above equations  can be solve in few steps. Firstly, we
introduce the following "relative" position variables
\begin{equation}
\vec{\hat{x}}_{a\,{\rm com}}=\vec{x}_{a+1\,{\rm
com}}-\vec{x}_{a\,{\rm com}}\;\;\;;\;\;\; a=1,\,2,\dots,\,
N-1\;,\label{szukalski2034}
\end{equation}
and then, we have
\begin{equation}
x_{n\,{\rm com}}^i=\sum_{a=1}^{n-1}\hat{x}_{a\,{\rm
com}}^i+x_{1\,{\rm com}}^i\;,\label{szukalski2035}
\end{equation}
for  $2\leq n\leq N$.\\
Next, we rewrite the Newton equation (\ref{szukalski201}) in terms
of variables (\ref{szukalski2034}) as follows
\begin{eqnarray}
\ddot{\vec{x}}_{1\,{\rm com}} &=&-\omega^2\vec{x}_{1\,{\rm
com}}+\omega M\,T\,\dot{\vec{x}}_{1\,{\rm com}}+
\vec{h}\;,\label{szukalski2036}\\
\ddot{\vec{\hat{x}}}_{a\,{\rm com}}
&=&-\left(\omega^2+\frac{N\lambda^2}{m}\right)\vec{\hat{x}}_{a\,{\rm
com}}\;,\label{szukalski2037}
\end{eqnarray}
where
\begin{eqnarray}
&&M=N\,m\;\;\;,\;\;\;
T^{ij}=\omega\sum_{k=1}^3\theta_k\epsilon^{kij}\;,\label{swarzyca100000}\\
&&h^i=\frac{1}{m}\sum_{b=2}^{N}\sum_{a=1}^{b-1}\left(\lambda^2
\hat{x}_{a\,{\rm com}}^i+m^2 \omega \sum_{j=1}^3 T^{ij}
\dot{\hat{x}}_{a\,{\rm com}}^j\right)\;.\label{swarzyca100}
\end{eqnarray}
 The solution of equation (\ref{szukalski2037}) can be found easily,  and
 it takes the form
\begin{eqnarray}
\hat{x}_{a\,{\rm com}}^i(t)&=&\hat{x}_{a\,{\rm com}}^i(0)\cos(\Omega
t)+\frac{\hat{v}_{a\,{\rm com}}^i(0)}{\Omega}\sin(\Omega
t)\;\;\;;\;\;\;
\Omega=\sqrt{\omega^2+\frac{N\lambda^2}{m}}\;,\label{swarzyca101}\\
h^i(t)&=&h_s^i\sin(\Omega t)+h_c^i\cos(\Omega
t)\;,\label{swarzyca101}
\end{eqnarray}
with
\begin{eqnarray}
h_s^i&=&\frac{1}{\Omega
m}\sum_{b=2}^N\sum_{a=1}^{b-1}\left(\lambda^2\hat{v}_{a\,{\rm
com}}^i(0)-m^2\omega\Omega^2
\sum_{j=1}^3 T^{ij}\hat{x}_{a\,{\rm com}}^j(0) \right)\;,\label{swarzyca102}\\
h_c^i&=&\frac{1}{m}\sum_{b=2}^N\sum_{a=1}^{b-1}\left(\lambda^2\hat{x}_{a\,{\rm
com}}^i(0)+m^2\omega \sum_{j=1}^3 T^{ij}\hat{v}_{a\,{\rm com}}^j(0)
\right)\;,\label{swarzyca103}
\end{eqnarray}
and with symbols $\hat{x}_{a\,{\rm com}}^i(0)$, $\hat{v}_{a\,{\rm
com}}^i(0)$ denoting the initial "relative" positions and velocities
respectively. \\
Further, using the formula (\ref{swarzyca101}) we get the following
explicite  solution of the equation (\ref{szukalski2036})
\begin{eqnarray}
x_{1\,{\rm
com}}^1(t)&=&\frac{1}{\omega\sqrt{4+M^2\omega^2\theta^2}}\left((v_{1\,{\rm
com}}^1(0)+\Omega_3 x_{1\,{\rm com}}^2(0))\sin(\Omega_2 t)\right.\nonumber\\
&~&~~~~~~~~~~~~~~~~~~~-~\;(v_{1\,{\rm com}}^2(0)-\Omega_3 x_{1\,{\rm
com}}^1(0))
\cos(\Omega_2 t))\;+\nonumber\\
&+&\left. (v_{1\,{\rm com}}^1(0)-\Omega_2 x_{1\,{\rm
com}}^2(0))\sin(\Omega_3 t)+(v_{1\,{\rm com}}^2(0)+\Omega_2
x_{1\,{\rm com}}^1(0))\cos(\Omega_3 t)\right)\nonumber\\
&+&\frac{1}{(\Omega_2^2-\Omega^2)\sqrt{4+M^2\omega^2\theta^2}\omega}\left((h_s^2\Omega-h_c^1\Omega_2)
(\cos(\Omega_2 t)-
\cos(\Omega t))\right.\nonumber\\
&-&\left.(h_c^2\Omega_2+h_s^1\Omega)\sin(\Omega_2 t)+(h_s^1\Omega_2+h_c^2\Omega)\sin(\Omega t)\right)\nonumber\\
&+&\frac{1}{(\Omega_3^2-\Omega^2)\sqrt{4+M^2\omega^2\theta^2}\omega}\left(-(h_s^2\Omega+h_c^1\Omega_3)
(\cos(\Omega_3 t)-
\cos(\Omega t))\right.\nonumber\\
&+&\left.(h_c^2\Omega_3+h_s^1\Omega)\sin(\Omega_3
t)+(h_s^1\Omega_3-h_c^2\Omega)\sin(\Omega t)\right)\;,\\
 x_{1\,{\rm
com}}^2(t)&=&\frac{1}{\omega\sqrt{4+M^2\omega^2\theta^2}}\left((v_{1\,{\rm
com}}^1(0)+\Omega_3 x_{1\,{\rm com}}^2(0))\cos(\Omega_2 t)\right.\nonumber\\
&~&~~~~~~~~~~~~~~~~~~~ +\;(v_{1\,{\rm com}}^2(0)-\Omega_3 x_{1\,{\rm
com}}^1(0)) \sin(\Omega_2 t))\;+\nonumber\\
 &-&\left. (v_{1\,{\rm
com}}^1(0)-\Omega_2 x_{1\,{\rm com}}^2(0))\cos(\Omega_3
t)+(v_{1\,{\rm com}}^2(0)+ \Omega_2 x_{1\,{\rm
com}}^1(0))\sin(\Omega_3 t)\right)\nonumber\\
&+&\frac{1}{(\Omega_2^2-\Omega^2)\sqrt{4+M^2\omega^2\theta^2}\omega}\left(-(h_s^1\Omega+h_c^2\Omega_2)
(\cos(\Omega_2 t)- \cos(\Omega t))\right.\nonumber\\
&+&\left.(h_c^1\Omega_2-h_s^2\Omega)\sin(\Omega_2
t)+(h_s^2\Omega_2-h_c^1\Omega)\sin(\Omega t)\right)\nonumber\\
&+&\frac{1}{(\Omega_3^2-\Omega^2)\sqrt{4+M^2\omega^2\theta^2}\omega}\left((h_s^1\Omega-h_c^2\Omega_3)
(\cos(\Omega_3 t)- \cos(\Omega t))\right.\nonumber
\end{eqnarray}
\begin{eqnarray}
&-&\left.(h_c^1\Omega_3+h_s^2\Omega)\sin(\Omega_3 t)+(h_s^2\Omega_3+h_c^1\Omega)\sin(\Omega t)\right)\nonumber\;,\\
x_{1\,{\rm com}}^3(t)&=&\frac{1}{\omega}v_{1\,{\rm com}}^3(0)\sin(\omega t)+x_{1\,{\rm com}}^3(0)\cos(\omega t)\nonumber\\
&-&\frac{1}{\omega(\omega^2-\Omega^2)}\left(h_c^3\omega(\cos(\omega
t)-\cos(\Omega t))+h_s^3(\Omega \sin(\omega t)-\omega \sin(\Omega
t)) \right)\nonumber\;,
\end{eqnarray}
with the   assumption  $\vec{\theta} = [\,0,0,\theta\,]$, symbols
${x}_{1\,{\rm com}}^i(0)$ and ${v}_{1\,{\rm com}}^i(0)$ denoting the
initial  data for first particle, and $\Omega_2$, $\Omega_3$ given
by
\begin{equation}
\Omega_2 =\frac{\omega}{2}\left(\omega M\,\theta  + \sqrt{4
+\omega^2 M^2\,\theta^2} \right)\;\;\;,\;\;\; \Omega_3
=\frac{\omega}{2}\left(\sqrt{4 +\omega^2 M^2\,\theta^2}- \omega
M\,\theta\right)\;.
\end{equation}
The remaining trajectories $\vec{x}_{2\,{\rm
com}}(t),\,\ldots,\,\vec{x}_{N\,{\rm com}}(t)$ can be easily
obtained by use of
the formulas (\ref{szukalski2035}).\\

\section{Final remarks}

In this article we construct the classical model of $N$
nonrelativistic particles moving in noncommutative space-time
(\ref{canamm}). The corresponding equation of motion for arbitrary
spherically symmetric potential (\ref{pot}) are provided. In
particular, there are analyzed two distinguished examples of such
systems - the set of $N$ coupled oscillators as well as the system
of $N$ particles moving in the presence of gravitational field
provided by massive point-like source.

It should be noted that the presented considerations can be extended
at least in two directions. First of all, one can consider the
multiparticle system associated with the Lie-algebraically deformed
space-time (\ref{betanoncomm1}). Secondly, one can quantize the
analyzed above (classical) model by introducing the Schr\"{o}dinger
equation
 defined on deformed space-time
(\ref{canamm}). The studies in these directions already started and
are in progress.

\section*{Acknowledgments}
The authors would like to thank J. Lukierski
for valuable discussions.\\
This paper has been financially supported by Ministry of Science and
Higher Education grant NN202318534.

\end{document}